%Paper: hep-th/9208079
%From: matsuo@rabbit.math.nagoya-u.ac.jp (Atsushi Matsuo)
%Date: Mon, 31 Aug 92 20:56:40 JST

\baselineskip=22pt
\magnification=\magstep1
\font\germ=eufm10

\def\QED{{\it \hfill Q.E.D.}}

\def\Z{{\bf Z}}
\def\sl{{\hbox{\germ sl}}\,}
\def\ge{{\hbox{\germ g}}\,}
\def\qqquad{\qquad\qquad}

\def\and{\hbox{ \  and \ }}
\def\parn{\par\noindent}
\def\medn{\medskip\medskip\noindent}
\def\med{\medskip\medskip}
\def\ds{\displaystyle }
\def\no{\leqno\ \ }
\def\hb{\hfill\break}

\def\vbn{\vfill\break\noindent}

{\rm \hfill
August 31, 1992}
\medskip
{\bf
\centerline{Free Field Representation of Quantum Affine Algebra
$U_q(\widehat{\sl}_2)$}}
\medskip\medskip
\centerline{Atsushi MATSUO}\med
\centerline{Department of Mathematics}\par\noindent
\centerline{Nagoya University, Nagoya 464-01, Japan}
\med\medn
\baselineskip=17pt plus 3pt
\def\F{{\cal F}}
\parn
{\bf Abstract.} \quad
A Fock representation of the quantum affine algebra $U_q(\widehat{\sl}_2)$
is constructed by three bosonic fields for an arbitrary level with the help
of the Drinfeld realization.
\vbn
\beginsection{1. Introduction}\par
Recently the theory of the q-deformed chiral vertex operators (qVO) is
developed by Frenkel and Reshetikhin [1] based on the representation theory
of the quantum affine algebra $U_q(\hat\ge)$ [2].
They derive a q-difference equation for the n-point correlation function,
which is a q-analoque of the Knizhnik-Zamolodchikov equation (KZ) in the
Wess-Zumino-Witten (WZW) model [3] and is called the quantum
Knizhnik-Zamolodchikov
equation (qKZ).
The importance of this model is due to the fact that some elliptic solutions
of the quantum Yang-Baxter equation (YBE) of face type, including the
ABF-solution [4], are obtained as the connection matrices of
qKZ\footnote{$^1$}{Aomoto et al.\ [5] have independently shown that the
ABF-solution
is obtained as the connection matrix of a q-difference equation,
which is nothing else but a special case of qKZ [6].}.
\par
For a detailed study of the model, a free field representation of $U_q(\hat
\ge)$
seems be an essential machinery.
Jimbo et al.\ [7] explicitly calculate qVO for level one
$U_q(\widehat{\sl}_2)$-modules
by making use of the free field representation obtained by Frenkel and Jing
[8],
which is a q-deformation of the Frenkel-Kac construction [10].
We note that the Drinfeld realization of $U_q(\hat \ge)$ [11] is the main
tool in their works.
However a free field representation for an arbitrary level has not been known
even in the $U_q(\widehat{\sl}_2)$ case.
\par
When $q=1$, the currents and the chiral vertex operators can be constructed by
some free fields in general.
In fact the Wakimoto representation [12] of the affine Lie algebra $\widehat
{\sl}_2$
is quite useful for the purpose.
It is realized by one set of a $\beta$-$\gamma$ ghost system and a free bosonic
field,
and the vertex operators and the screening operators are explicitly written
down [13].
This construction allows a generalization to the higher rank case [14],
see also [16].
Remarkably, in the $\sl_2$ case, the materials are also realized by three
bosonic fields [15].
For instance, the standard $\sl_2$ currents of level $k$ are expressed as
$$\matrix{J^\pm(z)=\,:{1\over{\sqrt{2}}}\left[\sqrt{k+2}\partial\phi_1(z)\pm i
\sqrt{k}\partial\phi_2(z)\right]e^{\pm\sqrt{{{2\over
k}}}[i\phi_2(z)-\phi_0(z)]}:,\cr
\cr J^0(z)=-\sqrt{\ds{k\over2}}\partial\phi_0(z),\cr}\no(1.1)$$
where $\phi_i(z)$ are independent bosonic fields normalized as
$\phi_i(z)\phi_i(w)\sim \log(z-w)$.
\par
In this letter we shall generalize (1.1) to the case of the
quantum affine algebra $U_q(\widehat{\sl}_2)$.
In section 2, we review the Drinfeld realization of $U_q(\widehat{\sl}_2)$,
whose generating functions play the role of the $\sl_2$ currents.
In section 3, we express them by means of three bosonic fields with slightly
modified normalizations and expressions.
Section 4 is devoted to a brief conclusion and discussion.
\par
While completing this work the author learned that J.\ Shiraishi has also
obtained
a free field realization of $U_q(\widehat {\sl}_2)$ for an arbitrary level in a
different expression [19].
\vbn

\beginsection{2. Drinfeld realization of $U_q(\widehat{\sl}_2)$}\par
\medn
Here and after we frequently use the notation
$$[m]={{q^m-q^{-m}}\over{q-q^{-1}}}.\no(2.1)$$
The quantum affine algebra $U_q(\widehat{\sl}_2)$ is isomorphic to the
associative
algebra generated by the letters
$\{x_m^{\pm}\,|\, m\in \Z\}$, $\{a_m\,|\,m\in \Z_{\ne0}\}$, $q^{\pm {c\over2}}$
and
$q^{\pm a_0}$, satisfying the following defining relations:
$$q^{\pm {c\over2}}\in {\hbox{the center of the algebra}},\no(2.2)$$
$$[a_m,a_n]=\delta_{m+n,0}{{[2m][mc]}\over{m}},\ [a_m,q^{a_0}]=0,\no(2.3)$$
$$q^{a_0}x_m^\pm q^{-a_0}=q^{\pm2}x_m^\pm,\ [a_m,x_n^\pm]=
\pm [2m]q^{\mp |m|c}x_{m+l}^\pm,\no(2.4)$$
$$x_{m+1}^\pm x_{n}^\pm -q^{\pm2}x_n^\pm x_{m+1}^\pm=
q^{\pm2} x_{m}^\pm x_{n+1}^\pm -x_{n+1}^\pm x_{m}^\pm,\no(2.5)$$
$$[x_m^+,x_n^-]={1\over{q-q^{-1}}}(q^{{{(m-n)}\over2}c}\psi_{m+n}-q^{{{(n-m)}\over2}c}
\varphi_{m+n}),\no(2.6)$$
where $q^{\pm mc}$ for $m>0$ is understood as $(q^{\pm{c\over2}})^{2m}$ and
$\{\psi_r,\varphi_s\,|\,r\in \Z_{\geq 0},\,s\in \Z_{\leq 0}\}$ are related to
$\{a_m\,|\,m\in\Z_{\ne0}\}$ by
$$\eqalign{
  \sum_{m=0}^\infty \psi_m z^{-m}&=q^{a_0}\exp\bigl((q-q^{-1})\sum_{m=1}^\infty
a_m z^{-m}\bigr),\cr
 \sum_{m=0}^\infty \varphi_{-m}
z^{m}&=q^{-a_0}\exp\bigl(-(q-q^{-1})\sum_{m=1}^\infty
a_{-m} z^{m}\bigr).\cr}\no(2.7)$$
\par
Now consider the following generating functions:
$$\matrix{k_+(z)=\sum_{m=0}^\infty \psi_m z^{-m},\ k_-(z)=
\sum_{m=0}^\infty \varphi_{-m} z^{m},\cr\cr
x^\pm(z)=\sum_{m\in\Z}x^\pm_m z^{-m}.\cr}\no(2.8)$$
Compositions of these operators are defined as a formal power series.
Suppose that they act on a highest weight module and that $q^{{c\over 2}}$ acts
by a scalar $q^{{k\over 2}}$ for some complex number $k$.
Then we may understand them to be analytically continued outside some locus,
and the operators (2.8) are characterized by the following properties:
$$[k_\pm(z),\,k_\pm(z)]=0,\ {{q^{k+2}z-w}\over{q^{k}z-q^2w}}k_-(z)k_+(w)=
k_+(w)k_-(z){{q^{-k+2}z-w}\over{q^{-k}z-q^2w}},\no(2.9)$$
$$\eqalign{&k_+(z)x^\pm(w)k_+(z)^{-1}=
\left({{q^{\mp {k\over2}+2}w-z}\over{q^{\mp
{k\over2}}w-q^2z}}\right)^{\mp1}x^\pm(w)\cr
&k_-(z)x^\pm(w)k_-(z)^{-1}=\left({{q^{\mp {k\over2}+2}z-w}\over
{q^{\mp {k\over2}}z-q^2w}}\right)^{\pm1}x^\pm(w),}\no(2.10)$$
$$(z-q^{\pm2}w)x^\pm(z)x^\pm(w)=(q^{\pm2}z-w)x^\pm(w)x^\pm(z)\and\no(2.11)$$
$$x^+(z)x^-(w)\sim{1\over{q-q^{-1}}}\left({z\over{z-q^{k}w}}k_+(q^{{k\over2}}w)-
{z\over{z-q^{-k}w}}k_-(q^{-{k\over2}}w)\right).\no(2.12)$$
The last formula (opertor product expansion) means that $x^+(z)x^-(w)$ is
analytically continued with the singular part being the right hand side and it
coincides with $x^-(w)x^+(z)$.
The defining relations (2.2)--(2.6) are recovered from these properties by a
standard
argument, see [9].
\par
When $q$ goes to $1$, the algebra $U_q(\widehat{\sl}_2)$ goes to the enveloping
algebra $U(\widehat {\sl}_2)$ of the affine Lie algebra $\widehat {\sl}_2$,
and the operators (2.8) go to the standard $\sl_2$ currents by the following
correspondence:
$$x^\pm(z)\rightarrow zJ^\pm(z),\ k^+(z)-k^-(z)\rightarrow 2zJ^0(z).\no(2.13)$$

\beginsection{3. Free field representation of $U_q(\widehat{\sl}_2)$}\par
Let $k$ be a complex number.
Let $\{\alpha_\sigma,\,\alpha_\sigma(n)\,|\,\sigma=\pm1,\ n\in\Z\}$ be
a set of operators satisfying the following relations:
$$[\alpha_\sigma(m),\,\alpha_\tau(n)]=\sigma\delta_{\sigma,\tau}
\delta_{m+n,0}{{[2m][km]}\over {m}},\qquad [\alpha_\sigma(m),\,\alpha_\tau]=
\sigma\delta_{\sigma,\tau}\delta_{m,0}.\no(3.1)$$
Here $\delta$ denotes the Kronecker symbol.
Let $\{\beta(n)\,|\,\ n\in\Z_{\ne 0}\}$ be another set of operators satisfying
$$[\beta(m),\,\beta(n)]=\delta_{m+n,0}{{[2m][(k+2)m]}\over {m}}\no(3.2)$$
Suppose that they commute with $\alpha_\sigma$ and $\alpha_\sigma(m)$ for any
$m$.
\par
The Fock space $\tilde\F$ on which these operators act is supposed to be
generated by the negative modes $\alpha_1(m)$, $\alpha_{-1}(m)$, $\beta(m)$,
for
$m<0$, and by $\alpha_1$ and $\alpha_{-1}$ acting on the vacuum vector $v$
satisfying the following conditions:
$$\alpha_1(m)v=\alpha_{-1}(m)v=\beta(m)v=0 {\hbox { \ for any $m>0$, \
and}}\no(3.3)$$
$$\alpha_1(0)v \and \ \alpha_{-1}(0)v {\hbox{ \ are scalar multiples of
$v$.}}\no(3.4)$$
\par
We set
$$\eqalign{
K_{+}(z)&=\exp\left\{(q-q^{-1})\sum_{m=1}^\infty z^{-m}\alpha_1(m)\right\}
q^{\alpha_1(0)},\cr
K_{-}(z)&=\exp\left\{-(q-q^{-1})\sum_{m=1}^\infty z^{m}\alpha_1(-m)\right\}
q^{-\alpha_1(0)},\cr}\no(3.5)$$
$$\eqalign{
X^{+}(z)&={1\over{q-q^{-1}}}\biggl\{Y^+(z)Z_+(q^{-{{k+2}\over2}}z)W_+(q^{-{k\over2}}z)\cr
&\qqquad\qqquad-W_-(q^{{k\over2}}z)Z_-(q^{{{k+2}\over2}}z)Y^+(z)\biggr\},\cr
X^{-}(z)&={-1\over{q-q^{-1}}}\biggl\{Y^-(z)Z_+(q^{{{k+2}\over2}}z)
W_+(q^{{k\over2}}z)^{-1}\cr&\qqquad\qqquad-W_-(q^{-{k\over2}}z)^{-1}
Z_-(q^{-{{k+2}\over2}}z)Y^-(z)\biggr\},\cr}\no(3.6)$$
where
$$\eqalign{
Y^{+}(z)
  &=\exp\left\{\sum_{m=1}^\infty q^{-{{km}\over 2}}{{z^m}\over{[km]}}
\bigl(\alpha_1(-m)+\alpha_{-1}(-m)\bigr)\right\}\cr
  &\qquad e^{2(\alpha_1+\alpha_{-1})}z^{{1\over
k}(\alpha_1(0)+\alpha_{-1}(0))}\exp
\left\{-\sum_{m=1}^\infty
q^{-{{km}\over2}}{{z^{-m}}\over{[km]}}\bigl(\alpha_1(m)+
\alpha_{-1}(m)\bigr)\right\},\cr
Y^{-}(z)
  &=\exp\left\{-\sum_{m=1}^\infty q^{{{km}\over 2}}{{z^m}\over{[km]}}
\bigl(\alpha_1(-m)+\alpha_{-1}(-m)\bigr)\right\}\cr
  &\qquad e^{-2(\alpha_1+\alpha_{-1})}z^{-{1\over
k}(\alpha_1(0)+\alpha_{-1}(0))}
\exp\left\{\sum_{m=1}^\infty
q^{{{km}\over2}}{{z^{-m}}\over{[km]}}\bigl(\alpha_1(m)+
\alpha_{-1}(m)\bigr)\right\},\cr}\no(3.7)$$
$$\eqalign{
Z_+(z)&=\exp\left\{-(q-q^{-1})\sum_{m=1}^\infty z^{-m}{{[m]}\over{[2m]}}
\alpha_{-1}(m)\right\}q^{-{1\over 2}\alpha_{-1}(0)},\cr
Z_-(z)&=\exp\left\{(q-q^{-1})\sum_{m=1}^\infty z^{m}{{[m]}\over{[2m]}}
\alpha_{-1}(-m)\right\}q^{{1\over 2}\alpha_{-1}(0)},\cr}\no(3.8)$$
$$\eqalign{
W_{+}(z)&=\exp\left\{-(q-q^{-1})\sum_{m=1}^\infty z^{-m}{{[m]}\over{[2m]}}
\beta(m)\right\},\cr
W_{-}(z)&=\exp\left\{(q-q^{-1})\sum_{m=1}^\infty z^{m}{{[m]}\over{[2m]}}
\beta(-m)\right\}.\cr}\no(3.9)$$
\medn
{\bf Proposition. } By analytic continuation, $X^\pm(z)$ and $K_\pm(z)$ satisfy
the same relations as (2.9)--(2.12).
\medn{\it Proof. } The relation (2.9) is obvious by the definition.
The proofs of (2.10)--(2.11) are straightforward by calculating commutators of
the fields.
For example, the first relation of (2.10) follows from the following:
$$\eqalign{
K_+(z)Y^\pm(w) &=q^{\pm2}\exp\left\{\pm\sum_{m=1}^\infty
z^{-m}w^mq^{-{{km}\over2}}
{{q^{2m}-q^{-2m}}\over m}\right\}Y^\pm(w)K_+(z) \cr
&=\left({{q^2z-q^{\mp{k\over2}}w}\over{z-q^{2\mp{k\over2}}w}}\right)^{\pm1}
Y^\pm(w)K_+(z) .\cr}$$
Here we have used the formula: $\sum_{m=1}^\infty {{x^m}\over m}=-\log(1-x)$.
To prove (2.12) we put $X^+(z)=\{A(z)-B(z)\}/(q-q^{-1})$ and
$X^-(z)=-\{C(z)-D(z)\}/(q-q^{-1})$ where
$$\matrix{\ds{A(z)=Y^+(z)Z_+(q^{-{{k+2}\over2}}z)W_+(q^{-{k\over2}}z),}&
\ds{B(z)=W_-(q^{{k\over2}}z)Z_-(q^{{{k+2}\over2}}z)Y^+(z),}\hfill\cr
\cr
\ds{C(z)=Y^-(z)Z_+(q^{{{k+2}\over2}}z)W_+(q^{{k\over2}}z)^{-1},}&
\ds{D(z)=W_-(q^{-{k\over2}}z)^{-1}Z_-(q^{-{{k+2}\over2}}z)Y^-(z).}\cr}$$
Then we have
$$\eqalign{A(z)C(w)
&={{q^{-1}z-q^{k+1}w}\over{z-q^{k}w}}Y^+(z)Y^-(w)Z_+(q^{-{{k+2}\over2}}z)
Z_+(q^{{{k+2}\over2}}w)W_+(q^{-{k\over2}}z)W_+(q^{{k\over2}}w)^{-1}\cr
&\sim-{{(q-q^{-1})z}\over{z-q^{k}w}}Y^+(q^k w)Y^-(w)Z_+(q^{{k \over 2}-1}w)
Z_+(q^{{k\over 2}+1}w)\cr
&=-{{(q-q^{-1})z}\over{z-q^{k}w}}K_+(q^{{k\over2}}w).\cr}$$
Similarly we have
$$B(z)D(w)\sim{{(q-q^{-1})z}\over{z-q^{-k}w}}K_-(q^{-{k\over2}}w),$$
and the other products $A(z)D(w)$ and $B(z)C(w)$ are regular.
The proof of the relation $[X^+(z),\,X^-(w)]=0$ is straightforward.
\QED
\med
Now suppose that the vacuum vector $v$ of the Fock space $\tilde\F$
satisfies the condition
$${1\over k}\bigl(\alpha_1(0)+\alpha_{-1}(0)\bigr)v=mv
{\hbox{ for some integer $m$.}}\no(3.10)$$
Consider the subspace $\F$ of $\tilde\F$ generated by the actions of
the negative modes and of $\alpha_1+\alpha_{-1}$ on $v$.
Then it is clear by the definition that the mode expansion of
$X^\pm(z)$ and $K_\pm(z)$ like (2.8) makes sense on $\F$ and that each
Fourier component acts there.
Thus we obtain a representation of the algebra $U_q(\widehat{\sl}_2)$ on
$\F$ with the level $c=k$.

\beginsection{4. Conclusion}\par
In this letter we have constructed the Fock representation of the
quantum affine algebra $U_q(\widehat{\sl}_2)$ for an arbitrary level $k$ in
terms of three bosonic fields.
In the $q\rightarrow 1$ limit our representation goes to the representation of
the affine Lie algebra $\widehat{\sl}_2$ defined by the mode expansion of
the currents (1.1).
It is equivalent to a bosonization of the Wakimoto representation by a
certain transformation [15].
\par
In [16] solutions to KZ are explicitly constructed in the context of the
Wakimoto representation, and they give rise to the integral solutions obtained
previously [17].
The author expects that the Jackson integral solutions of qKZ [6,17] would
be obtained in our formulation.
\par
The present work will contribute to a better understanding of massive
deformations of conformal field theory as the Wakimoto representation did
in the WZW model.
A detailed analysis of the representation and a construction of qVO and
the screening operators will be contained in a separate paper.

\beginsection{Acknowledgement.}\par
The author thanks K.\ Kimura, T.\ Miwa and J.\ Shiraishi for discussion.
\vbn
\beginsection{References.}\par
\medn
\item{[1]}  I.B.\ Frenkel and N.Yu.\ Reshetikhin, Commun.\ Math.\ Phys.\ 146
(1992) 1.
\item{[2]}  V.G.\ Drinfeld, Sov.\ Math.\ Dokl.\ 32 (1985) 254.\hb
M.\ Jimbo, Lett.\ Math.\ Phys.\ 10 (1985) 63.
\item{[3]}  V.G.\ Knizhnik and A.B.\ Zamolodchikov, Nucl.\ Phys.\ B247 (1984)
83.\hb
A.\ Tsuchiya and Y.\ Kanie, Adv.\ Stud.\ Pure.\ Math 16 (1988) 297.
\item{[4]}  G.E.\ Andrews, R.J.\ Baxter and P.J.\ Forrester, J.\ Stat.\ Phys.\
35 (1984) 193.
\item{[5]}  K.\ Aomoto, Y.\ Kato and K.\ Mimachi, Int.\ Math.\ Res.\ Notes
(Duke Math.\ J.) 65 (1992) 7.
\item{[6]}  A.\ Matsuo, Jackson integrals of Jordan-Pochhammer type and quantum
Knizhnik-Zamolodchikov equations, preprint (1991) to appear in Commun.\ Math.\
Phys.
\item{[7]} M.\ Jimbo, K.\ Miki, T.\ Miwa and A.\ Nakayashiki, preprint RIMS-873
(1992).
\item{[8]} I.B.\ Frenkel and N.H.\ Jing, Proc.\ Nat'l Acad.\ Sci.\ USA 85
(1988) 9373.
\item{[9]} D.\ Bernard, Lett.\ Math.\ Phys.\ 17 (1989) 239.
\item{[10]} I.B.\ Frenkel and V.G.\ Kac, Invent.\ Math.\ 62 (1980) 23.
\item{[11]} V.G.\ Drinfeld.\ Sov.\ Math.\ Dokl.\ 36 (1987) 212.
\item{[12]}  M.\ Wakimoto, Commun.\ Math.\ Phys.\ 104 (1986) 605.

\item{[13]} D.\ Bernard and G.\ Felder, Commun.\ Math.\ Phys.\ 127 (1990) 145.
\item{[14]} B.\ Feigin and E.\ Frenkel, Commun.\ Math.\ Phys.\ 128 (1990)
161.\hb
P.\ Bouwknegt, J.\ McCarythy and K.\ Pilch, Prog.\ Theor.\ Phys.\ Suppl.\ 102
(1990) 67.
\item{[15]}  D.\ Nemeschansky, Phys.\ Lett.\ B224 (1989) 121.\hb
K.\ Ito.\ Nucl.\ Phys.\ B332 (1990) 566.\hb
T.\ Jayaraman, K.S.\ Narain and M.H.\ Sarmadi, Nucl.\ Phys.\ B343 (1990) 418.
\item{[16]} G.\ Kuroki, Commun.\ Math.\ Phys.\ 142 (1991) 511.\hb
H.\ Awata, A.\ Tsuchiya and Y.\ Yamada, Nucl.\ Phys.\ B365 (1991) 680.
\item{[17]}
E.\ Date, M.\ Jimbo, A.\ Matsuo and T.\ Miwa, Intern.\ J.\ Mod.\ Phys.\ B4
(1990) 1049.\hb
A.\ Matsuo, Commun.\ Math.\ Phys.\ 134 (1990) 65.\hb
V.V.\ Schechtman and A.N.\ Varchenko,\ Invent.\ Math.\ 106 (1991) 139.
\item{[18]}  A.\ Matsuo, Quatum algebra structure of certain Jackson integrals,
preprint (1992).\hb
N.Yu.\ Reshetikhin, Jackson type integrals, Bethe vectors, and solutions to a
difference analog of the Knizhnik-Zamolodchikov system, preprint (1992).
\item{[19]} J.\ Shiraishi.\ Free boson representation of
$U_q(\widehat{\sl}_2)$,
in preparation.
\bye